# Exciton spin dynamics in spherical CdS quantum dots

P. Horodyská,[1] P. Němec,[1,*] D. Sprinzl,[1] P. Malý,[1] V. N. Gladilin,[2,†] and J. T. Devreese[2]

[1]*Faculty of Mathematics and Physics, Charles University in Prague, Ke Karlovu 3, 121 16 Prague 2, Czech Republic*
[2]*Theoretische Fysica van de Vaste Stoffen, Departement Fysica, Universiteit Antwerpen, Groenenborgerlaan 171, B-2020 Antwerpen, Belgium*

**ABSTRACT**

Exciton spin dynamics in quasi-spherical CdS quantum dots is studied in detail experimentally and theoretically. Exciton states are calculated using the 6-band **k·p** Hamiltonian. It is shown that for various sets of Luttinger parameters, when the wurtzite lattice crystal field splitting and Coulomb interaction between the electron-hole pair are taken into account exactly, both the electron and hole wavefunction in the lowest exciton state are of *S*-type. This rules out the spatial-symmetry-induced origin of the dark exciton in CdS quantum dots. The exciton bleaching dynamics is studied using time- and polarization-resolved transient absorption technique of ultrafast laser spectroscopy. Several samples with a different mean size of CdS quantum dots in different glass matrices were investigated. This enabled the separation of effects that are typical for one particular sample from those that are general for this type of material. The experimentally determined dependence of the electron spin relaxation rate on the radius of quantum dots agrees well with that computed theoretically.

## I. INTRODUCTION

Semiconductor quantum dots (or nanocrystals) are an interesting type of matter that represents a transition between bulk materials and molecular compounds. If the size of quantum dots (QDs) is comparable to the exciton Bohr radius, they exhibit optical properties that differ dramatically from those of the bulk materials. In particular, the spatial confinement of an electron-hole pair by the QD boundary leads to the quantization of valence and conduction bands, which results in discrete atomic-like transitions that shift to higher energies as the size of QDs decreases[1]. This effect enables to tune the wavelength of light emitted from QDs, which is very attractive for various applications including construction of lasers[2] and preparation of fluorescence tags in biotechnology applications[3]. Moreover, quantum dots provide the possibility to test various theoretical models in three-dimensionally confined systems. The electronic structure of QDs is determined by the relative importance of different terms: the confinement induced level splitting, the electron-hole Coulomb and exchange interactions, crystal field, and the spin-orbit coupling.

---

[*] corresponding author: nemec@karlov.mff.cuni.cz
[†] currently working at INPAC, Katholieke Universiteit Leuven, Celestijnenlaan 200D, B-3001 Leuven, Belgium

The most often studied spherical QDs are made of II-VI compounds with CdSe as a most thoroughly investigated example[1,4,5,6]. The theoretical treatment of CdSe is strongly simplified by the fact that the spin-orbit coupling is large compared to the other interactions (the spin-orbit splitting energy is 420 meV (Ref. 7) and, therefore, the mixing between the light-hole and heavy-hole bands, on the one hand, and the spin-orbit split-off bands, on the other hand, may be neglected[5]. The detailed theoretical analysis in the framework of 4-band **k·p** model of the band-edge exciton fine structure in CdSe revealed that the originally eightfold degenerate lowest $(1S_e, 1S_{3/2})$ exciton energy level is split by the crystal field and the electron-hole exchange interaction into five levels and that the lowest state is optically passive (dark exciton) (Ref. 5). On the other hand, in CdS, which has the same wurtzite hexagonal crystal structure, the SO-splitting is only 68 meV (Ref. 7). The mixing between light-hole, heavy-hole and spin-orbit split-off bands can considerably affect the exciton optical properties near the absorption edge[7] and may even alter the order of the calculated hole energy levels. In particular, the 4-band model in CdS gives the lowest hole state always of *S*-type (Refs. 8, 9) while in the 6-band **k·p** model the lowest hole state in small nanocrystals is of the *P*-symmetry[7,10,11,12,13]. As the lowest electron state has always an *S*-symmetry, it was believed that the lowest e-h pair state in small CdS QDs is an *orbital-symmetry-forbidden* dark exciton[11]. However, the situation changes when the excitonic effects (Coulomb and exchange interaction between the electron and hole) are taken into account. In Ref. 32 it was shown using the 6-band **k·p** model that in CdSe the lowest hole *P*-state shifts due to the excitonic effects less than the above lying *S*-state, thus the region of the QD sizes, where the *S*-state is the lowest hole state, enlarges. Similar result was obtained also for CdS QDs (Ref. 12), but in this paper a cubic zinc-blende crystal structure was considered (i.e., the effect of the crystal field was neglected). Another problem in the 6-band modeling is that the results are extremely sensitive to the values of effective mass parameters as was shown in Refs. 7, 5, 13, 9 (even in CdSe for one set of parameters the lowest hole state is of *P*-type - see Fig. 2 in Ref. 7). As the Luttinger material parameters, needed in the effective mass approximation (EMA), are not exactly known for CdS, the computed lowest state is sometimes optically dark and sometimes bright[9].

Understanding of the exciton fine structure is also important because it is directly connected with the spin relaxation measurements by the polarization-sensitive methods of ultrafast laser spectroscopy[14]. Spin dynamics is of fundamental importance in many spintronic applications because it determines the survival of information encoded in the carrier spin[15]. In general, QDs are interesting for spintronics for their long spin coherence times that is a consequence of the suppression of the spin relaxation channels, which are dominant in bulk materials[16]. Recently, we have shown that the spin relaxation is even more suppressed in spherical QDs than in the highly anisotropic self-assembled QDs (Refs. 16, 17, 18) and that the corresponding relaxation times can reach nanosecond values even at room temperature[14].

In this paper, we report on a detailed comparison of the theoretical computation of the exciton fine structure with the experimental results obtained by the polarization-sensitive pump-probe technique of the ultrafast laser spectroscopy in quasi-spherical CdS QDs embedded in a glass matrix. The paper is organized as follows. In Sec. II we provide details about the investigated samples and about the experimental technique that was used to study the exciton dynamics. In Sec. III we show the exciton fine structure that was calculated for spherical CdS QDs taking into account the Coulomb and exchange interactions between the electron-hole pair as well as the wurtzite lattice crystal field splitting. In Sec. IV we present the obtained experimental results and we compare them thoroughly with the predictions of our model. The main conclusions are summarized in Sec. V.

## II. EXPERIMENTAL

In our time-resolved experiments, we studied CdS QDs in a glass matrix, which is a well known model material of mutually isolated quasi-spherical QDs (Ref. 19). The samples studied were platelets (with a thickness of about 200 μm) prepared by mechanical polishing from as-received color filter glasses made by Hoya (samples Y44 and Y46), Schott (sample GG435) and Corning (sample 3-73), which contain CdS QDs with wurtzite lattice and volume filling factor of about 0.1%. These particular samples were selected because they contain QDs with similar sizes but the host glass has considerably different chemical composition in the samples made by different producers[20]. In this way, we tried to separate the effects that are typical of one particular sample from those that are general for this type of material. The samples were characterized by their linear absorption spectra and by transmission electron microscopy (TEM). TEM analysis revealed that the majority of QDs in the samples is spherical or only slightly elliptical [typically, 70% QDs have $\varepsilon = d_{max}/d_{min} < 1.25$ (Ref. 21)] and their size distribution can be fitted well by a lognormal function with a width around 10% (Ref. 21). This slightly asymmetric shape of the size distribution with a tail extending to the larger sizes is in agreement with the results published for similar samples[19,22,23], which shows that for nanocrystals in a glass matrix the size distribution is Gaussian at early stages of the growth but it changes to an asymmetric distribution during the growth. The elongation of nanocrystals was suggested to be in a direction of the hexagonal c-axis[4,24], but we were not able to verify this for our nanocrystals. Therefore, in our model we assume that the nanocrystals have no preferential orientation. The carrier dynamics were investigated by the time-resolved pump & probe technique using titanium sapphire laser (Tsunami, Spectra Physics) with BBO frequency doubler that generates femtosecond pulses (time width of 80 fs, spectral width of 10 meV) with a repetition rate of 82 MHz. As a measure of transmission changes, we used the differential transmission $\Delta T/T_0 = (T_E - T_0)/T_0$, where $T_E$ ($T_0$) is the sample transmission measured by probe pulses with (without) the pump pulses. To exclude many-particle processes, all the measurements were done in the low fluence regime, where on average much less than one electron-hole pair per QD was excited. Probe pulses were 10 times weaker than pump pulses and they have the same photon energy (degenerate experiment). The polarization of the laser pulses was controlled using zero-order quarter wave plates and/or photo-elastic modulator (PEM). The measurements were done at temperatures from 8 K to 300 K, the sample was mounted on the cold finger of the closed-cycle helium cryostat (Janis Cryogenics).

A rate of the carrier energy relaxation and recombination was determined from the decay of the time-resolved polarization-insensitive dynamics of $\Delta T/T_0$, which is the average of the signals measured using probe pulses with the same and opposite circular polarization with respect to the circular polarization of the pump pulses[14]. The rate of the spin relaxation was determined from the degree of circular polarization of the measured signal $P_C$, which is defined as

$$P_C = \frac{\Delta T^{++} - \Delta T^{+-}}{\Delta T^{++} + \Delta T^{+-}}, \tag{1a}$$

where $\Delta T^{++}$ ($\Delta T^{+-}$) denote the signal measured with probe pulses with the same (opposite) circular polarization with respect to the circular polarization of pump pulses. Similarly, the degree of linear polarization of the measured signal $P_L$ is defined as

$$P_L = \frac{\Delta T^{\parallel} - \Delta T^{\perp}}{\Delta T^{\parallel} + \Delta T^{\perp}}, \tag{1b}$$

where $\Delta T^{\parallel}$ ($\Delta T^{\perp}$) denote the signal measured with the co-linearly (cross-linearly) polarized probe pulses with respect to the linear polarization of pump pulses.

### III. THEORETICAL MODEL

#### A. Electron and hole energy levels

Different authors make different approximations in the calculations of the CdS QDs band structure. For example, in Ref. 9 the exciton effects (Coulomb and exchange interaction) are neglected, in Ref. 7 the Coulomb interaction is treated to the first order of perturbation, and in Ref. 12 the Coulomb interaction between the electron and hole is incorporated but the wurtzite crystal field splitting is neglected (i.e. the calculations are done for the cubic zinc-blende crystal structure). In conclusion, there seems to be a discrepancy in the literature about the band structure of spherical CdS QDs. Therefore, we find it useful to present here in more detail our exciton energy calculations for spherical CdS QDs with hexagonal lattice, where the effects of the crystal field, electron-hole Coulomb and exchange interaction are taken into account exactly.

There are several factors affecting the exciton states in a QD: the confining potential of the QD boundary, the crystal field of the lattice structure, the spin-orbit interaction and the Coulomb and exchange interaction between the electron and the hole. The corresponding Hamiltonian can be expressed as (see, e.g., Ref. 6 and references therein)

$$H_{ex} = \frac{1}{2m_e}\mathbf{p}_e^2 + H_h + V_C(\vec{r}_e,\vec{r}_h) + V_{exch}(\vec{r}_e,\vec{r}_h) + V_{hex} \tag{2}$$

where the first term is the kinetic energy of the electron, while the second term is the Hamiltonian of a hole in cubic lattice. $V_C$ is the Coulomb potential between the electron and hole, $V_{exch}$ is the electron-hole exchange interaction, and the last term $V_{hex}$ describes the effect of the crystal field in a hexagonal lattice on hole states. These additional interactions are often treated as small perturbations like it was done, e.g., in Refs. 25 and 4 or in Ref. 7 for both CdSe and CdS QDs. Strictly speaking, such a perturbative approach is applicable only within a limited range of parameters (e.g., in the case, where the interlevel energy spacings in a QD are sufficiently large as compared to the interaction energies). Our approach, which we describe here, avoids these limitations. First, the so-called *bare* electron-hole states are calculated. These are the eigenstates of the Hamiltonian, which includes only the kinetic energy and the spin-orbit interaction for holes:

$$H_{eh} = \frac{1}{2m_e}\mathbf{p}_e^2 + H_h . \tag{3}$$

Second, the whole Hamiltonian (2) is diagonalized numerically in the basis of the 100-500 lowest *bare* states of the electron-hole pair. In the case of QDs embedded in the glass matrix, the boundaries with an infinite potential barrier can be safely assumed.

The wavefunctions of an electron confined in a spherically symmetric potential are given in the envelope function approximation by

$$\Psi^{(e)}_{n_e l m s_z}(\vec{r}) = \sqrt{\frac{2}{R^3}} \cdot \frac{J_l\left(\chi_{n_e l} \frac{r}{R}\right)}{J_{l+1}(\chi_{n_e l})} |l,m\rangle \left|\frac{1}{2}, s_z\right\rangle \quad (4)$$

where $|l,m\rangle$ is the eigenfunction of the angular momentum operator of the envelope function, $\left|\frac{1}{2}, s_z\right\rangle$ with $s_z = \pm 1/2$ is the Bloch wavefunction at the bottom of the $S$-type conduction band, $R$ is the QD radius and $\chi_{nl}$ is the $n^{th}$ zero of the $l^{th}$ Bessel function $J_l(\chi)$. The index $n_e$ labels different radial solutions in order of increasing energy. Due to strong size quantization for conduction electrons, only the lowest electron states, 1$S$, are relevant to optical properties of QDs near the absorption edge.

A hole confined to a spherical QD possesses a definite eigenvalue of the *total* hole angular momentum **F**, which is a sum of the hole spin **J** ($J = 1/2, 3/2$) and the nonzero orbital angular momentum of the envelope function **L**: $\mathbf{F} = \mathbf{L} + \mathbf{J}$. The Luttinger-Kohn Hamiltonian $H_h$, taken in so-called spherical approximation where one neglects the warping of the valence band connected with the cubic symmetry of the semiconductor lattice (see, e.g., Refs. 26 and 30), couples states with the same $F$. Therefore, hole states in a spherical QD are – in general – superpositions of components with different $L$: $L = F \pm 1/2, F \pm 3/2$. Since the Hamiltonian $H_h$ preserves the symmetry of states with respect to inversion of co-ordinates (i.e. only states with the same parity of the quantum number $L$ can be mixed), the difference in $L$ between the mixed components is $\Delta L = 0, 2$ (Refs. 22, 23, 26, 30). In view of relatively small spin-orbit splitting in CdS, a 6-band **k·p** model is used to describe the hole states. The *bare* hole states can be expressed[26] as

$$\Psi^{(h)}_{n_h F F_z q}(\mathbf{r}) = R^{(1)}_{n_h F q}(r)\left|F+\frac{q}{2},\frac{3}{2},F,F_z\right\rangle + R^{(2)}_{n_h F q}(r)\left|F-\frac{3q}{2},\frac{3}{2},F,F_z\right\rangle + R^{(3)}_{n_h F q}(r)\left|F+\frac{q}{2},\frac{1}{2},F,F_z\right\rangle, \quad (5)$$

where $|L, J, F, F_z\rangle$ is the eigenfunction of the total angular momentum $F$ with the projection $F_z$ on the c-axis of the wurtzite crystal structure, $n_h$ labels the different radial solutions in order of increasing energy, and $q = \pm 1$. The functions $R^{(1)}_{n_h F q}(r)$ and $R^{(2)}_{n_h F q}(r)$ are the amplitudes of the components, corresponding to heavy and light holes, while the amplitude $R^{(3)}_{n_h F q}(r)$ is related to the split-off hole component. The eigenenergies must be calculated numerically.

Putting together the electron and hole states without their mutual interaction, the *bare* states of an electron-hole pair are obtained, which form the basis for the diagonalization of the Hamiltonian (3):

$$\Psi^{(eh)}_{n_e lms_z n_h FF_z q}(\mathbf{r}_e, \mathbf{r}_h) = \Psi^{(e)}_{s_z}(\mathbf{r}_e)\Psi^{(h)}_{n_h FF_z q}(\mathbf{r}_h). \tag{6}$$

For better readability, the electron-hole *bare* states will be denoted in a same way like in e.g. Ref. 6 as $(1S, nX_F)_{s_z F_z}$, where the first term stands for the state of the electron [note that we only take into account the lowest state with $n_e = 1$ and $l = 0$ (*S*-state)] and the second term denotes the $n^{th}$ radial state of a hole with the total angular momentum $F$; the envelope-function orbital states of the hole are labeled by $X = S$, $P$, $D$,... for $L = 0, 1, 2,...$, where $L$ is the *lowest* value of the envelope-function angular momentum in a superposition, which describes the corresponding hole state.

When compared to the four-band description for the hole states, one of the most important distinctive features of the results of the six-band model is that the energy spectra, given by the six-band model, are not merely scaled with the QD radius $R$ as $R^{-2}$. In particular, the order of states with different symmetry appears dependent on the QD size. While within the four-band model the lowest energy level of a hole in a CdS spherical QD is always the $1S_{3/2}$ state (not shown here, see e.g. Ref. 9), the application of the six-band model gives for small QDs the lowest energy level $1P_{3/2}$.

### B. Effect of the hexagonal lattice

In a spherical QD with cubic lattice, the hole states $nX_F$ with different *z*-projections of the total angular momentum of the hole, $F_z$, are degenerate. However, in a QD with wurtzite lattice, like CdS, the crystal field splits the aforementioned degenerate levels into sublevels with definite values of $|F_z|$, where $F_z$ is the projection of the total angular momentum of a hole on the *c*-axis of the wurtzite lattice. Like in Refs. 25 and 6, the effect of the crystal field in a hexagonal lattice on hole states is described here by the potential

$$V_{hex} = -\frac{\Delta}{2}\left(\hat{J}_z^2 - \frac{1}{4}\right) \tag{7}$$

where for CdS $\Delta = 27$ meV (Ref. 27). For the matrix elements of $V$ in the basis of hole wavefunctions (5), we obtain

$$\begin{aligned}
V_{nFF_z q;n'F'F'_z q'} = \delta_{F_z F'_z} &\{[\delta_{F+q/2, F'+q'/2} B^{(11)}_{nFq;n'F'q'} + \delta_{F+q/2, F'-3q'/2} B^{(12)}_{nFq;n'F'q'}] \\
&\times [C^{(+)}_{F,F',F+q/2,F_z} + C^{(-)}_{F,F',F+q/2,F_z}] \\
&+ [\delta_{F-3q/2, F'-3q'/2} B^{(22)}_{nFq;n'F'q'} + \delta_{F-3q/2, F'+q'/2} B^{(21)}_{nFq;n'F'q'}] \\
&\times [C^{(+)}_{F,F',F-3q/2,F_z} + C^{(-)}_{F,F',F-3q/2,F_z}]\},
\end{aligned} \tag{8}$$

where

$$B^{(jk)}_{nFq;n'F'q'} = \int_0^\infty dr\, r^2 R^{(j)}_{nFq}(r) R^{(k)}_{n'F'q'}(r), \tag{9}$$

$$C^{(\pm)}_{F,F',L,F_z} = \left(L, F_z \pm \frac{3}{2}, \frac{3}{2}, \mp\frac{3}{2}\bigg|L, \frac{3}{2}, F, F_z\right)\left(L, F_z \pm \frac{3}{2}, \frac{3}{2}, \mp\frac{3}{2}\bigg|L, \frac{3}{2}, F', F_z\right). \tag{10}$$

The hole states including the crystal field splitting are shown in Fig. 1. For the $1S_{3/2}$ state the splitting, induced by the crystal field, is larger than that for the $1P_{3/2}$ state, so that in a relatively wide range of the QD radius $R$ the lowest $1S_{3/2}$ sublevel appears very close to the $1P_{3/2}$ sublevels. Moreover, as distinct from the case of no crystal-field effect (not shown here), now the lowest hole state is of the $1P_{3/2}$ type only for $R < 3$ nm as seen in Fig. 1. It is worth mentioning that – although the potential (7) is not diagonal in the basis (5) – the crystal-field induced mixing of different states $nX_F$ with the same $F_z$ is almost negligible for QDs with radii $R \sim 2$ nm, which are relevant for our experiment. So, the hole states still can be adequately characterized by labels $nX_F$ and the value of $F_z$.

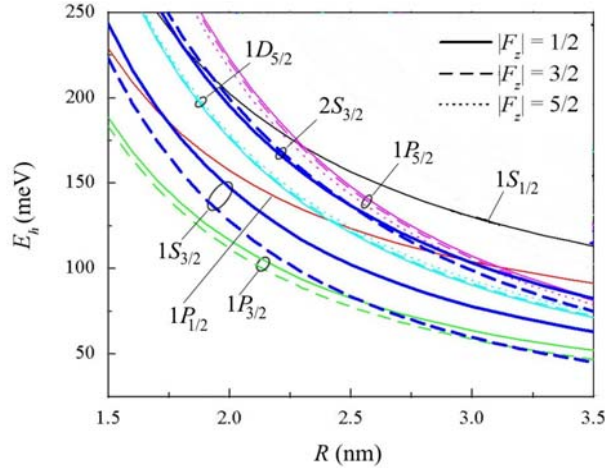

FIG. 1: Lowest energy levels of a hole in CdS spherical QDs of different radii $R$ calculated using the 6-band model with Luttinger parameters $\gamma_1 = 1.71$, $\gamma_2 = 0.62$ and the spin-orbit splitting $\Delta_{SO} = 70$ meV (Ref. 29) including the crystal field splitting.

### C. Excitonic effects

The Coulomb interaction between an electron and a hole in a spherical QD of radius $R$ is described by the potential[6,28]

$$V_C(\mathbf{r}_e, \mathbf{r}_h) = -\frac{e^2}{4\pi\varepsilon_0\varepsilon(\infty)} \sum_{l=0}^{\infty} P_l(\cos\vartheta_{eh}) \left\{ \frac{r_e^l}{r_h^{l+1}} \theta(r_h - r_e) + \frac{r_h^l}{r_e^{l+1}} \theta(r_e - r_h) \right.$$
$$\left. + \frac{(r_e r_h)^l}{R^{2l+1}} \frac{[\varepsilon(\infty) - \tilde{\varepsilon}(\infty)](l+1)}{\varepsilon(\infty)l + \tilde{\varepsilon}(\infty)(l+1)} \right\}, \quad (11)$$

where $e$ is the elementary charge, $\varepsilon_0$ is the permittivity of the vacuum, $P_l(x)$ is a Legendre polynomial of degree $l$, $\vartheta_{eh}$ is the angle between the electron and hole radius-vectors $\mathbf{r}_e$ and $\mathbf{r}_h$, $\varepsilon(\infty)$ and $\tilde{\varepsilon}(\infty)$ are the optical dielectric constants of a QD and its surrounding medium, respectively, $\theta(x) = 1$ for $x \geq 0$ and $\theta(x) = 0$ for $x < 0$. The physical reason for this relatively complicated form of the Coulomb interaction term is the difference in dielectric constants between the quantum dot material and the surrounding medium. This expression follows when expanding the interaction potential in spherical harmonics and then solving the corresponding Poisson equation with standard Maxwell's boundary conditions on the spherical surface of a quantum dot. An important point is that the Coulomb interaction does not mix states $(1S, nX_F)_{s_z F_z}$ with different $X$, $F$, $s_z$ or $F_z$. There is, however, a mixing between states with the same quantum numbers $X$, $F$, $s_z$, $F_z$ but different $n$. Nevertheless, for convenience, we denote each *exciton* state as $(1S, nX_F)_{s_z F_z}$, i.e., by a set of quantum numbers, which characterize the corresponding "parental" state, related to a non-interacting electron-hole pair. The actual meaning of these notations is that under a gradual decrease of the Coulomb interaction the exciton state, labeled as $(1S, nX_F)_{s_z F_z}$, smoothly evolves into the *bare* state of the electron-hole pair, where the electron is in the state 1S with the spin projection $s_z$ and the hole is in the state $nX_F$ with the z-projection of the hole angular momentum $F_z$. Basically, just the afore described "parental" bare state dominates the superposition, which describes the exciton state denoted as $(1S, nX_F)_{s_z F_z}$.

In Fig. 2 (a), the calculated energy spectrum of an *exciton* in a CdS spherical QD is shown as a function of the QD radius. It is noteworthy that downwards shift of the $(1S, 1S_{3/2})_{s_z F_z}$ exciton states due to the Coulomb interaction, as compared to the *bare* states $(1S, 1S_{3/2})_{s_z F_z}$ of a non-interacting electron-hole pair, is significantly larger than that for the $(1S, 1P_{3/2})_{s_z F_z}$ states. As a result, for the *exciton*, the lowest states appear to be $(1S, 1S_{3/2})_{s_z F_z}$ for all relevant QD sizes.

It should be mentioned that the calculated energy spectra are very sensitive to the values of the effective-mass parameters $\gamma_1$, $\gamma_2$, and $\Delta_{so}$, which are known for CdS with relatively large uncertainties. We performed the calculations with three different sets of effective-mass parameters: $\gamma_1 = 1.71$, $\gamma_2 = 0.62$, $\Delta_{so} = 70$ meV (Ref. 29); $\gamma_1 = 1.02$, $\gamma_2 = 0.41$, $\Delta_{so} = 62.4$ meV (Ref. 30); $\gamma_1 = 1.09$, $\gamma_2 = 0.34$, $\Delta_{so} = 68$ meV (Ref. 7). The obtained results are shown in Fig. 2(a)-(c). It is apparent that even though the exact order of the higher states and the energy distance between the individual states depend strongly on the values of the material parameters, the lowest state is always $(1S, 1S_{3/2})_{s_z F_z}$. In the following, unless explicitly noted differently, we will present results computed with the parameter set reported in Ref. 7, which seems to provide the best agreement between the theoretical and experimental results (see below).

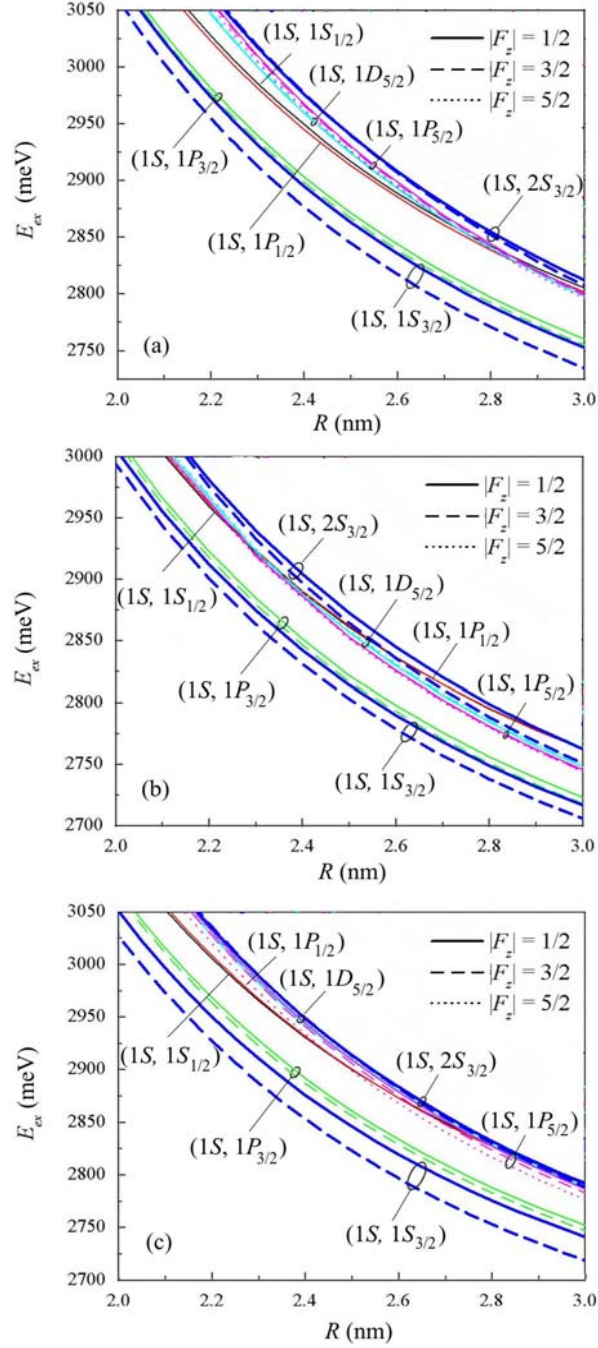

FIG. 2: Lowest energy levels of an exciton in CdS spherical QDs of different radii $R$. Calculations are performed using the 6-band model for holes with parameters (a) $\gamma_1 = 1.71$, $\gamma_2 = 0.62$, $\Delta_{SO} = 70$ meV (Ref. 29), (b) $\gamma_1 = 1.02$, $\gamma_2 = 0.41$, $\Delta_{SO} = 62.4$ meV (Ref. 30), and (c) $\gamma_1 = 1.09$, $\gamma_2 = 0.34$, $\Delta_{SO} = 68$ meV (Ref. 7).

Besides the Coulomb interaction, described by Eq. (11), there is also the exchange electron-hole interaction. The corresponding Hamiltonian can be written in the form[5]

$$V_{\text{exch}}(\mathbf{r}_e, \mathbf{r}_h) = -\frac{4}{3}\varepsilon_{\text{exch}} a_0^3 \delta(\mathbf{r}_e - \mathbf{r}_h)(\hat{\mathbf{s}} \cdot \hat{\mathbf{J}}), \qquad (12)$$

where $\hat{\mathbf{s}}$ and $\hat{\mathbf{J}}$ are the spin operators of the electron and the hole, respectively, $a_0$ is the lattice constant, $\varepsilon_{exch}$ is the exchange strength constant. Using Eq. (13) from Ref. 5 and the value $\hbar\omega_{ST} = 0.2$ meV for the exchange splitting in bulk CdS with wurtzite structure[31] one finds $\varepsilon_{exch} \approx 35$ meV for CdS.

The electron-hole exchange interaction couples the electron and hole spins. Due to the exchange interaction, the exciton energy levels with definite $|F_z|$ (which are degenerate with respect to $s_z$ and the sign of $F_z$ in the absence of the exchange interaction) are split into sublevels with definite $|N_z|$, where $N_z$ is the z-projection of the *total* angular momentum of the *exciton*, $N_z = s_z + F_z$. In Fig. 3 we show the exchange splitting for the lowest exciton state $(1S, 1S_{3/2})_{s_z,F_z}$. The two groups of levels are clearly separated in energy due to the crystal field induced splitting (when the exchange interaction is small enough, the lower group comes mainly from the state with $|F_z| = 3/2$ while the upper group has a dominant character of $|F_z| = 1/2$). The lower group consists of the dipole inactive twofold degenerate level $|N_z| = 2$ and dipole active twofold degenerate level with $|N_z| = 1$, denoted as $1^L$. The upper group consists of the dipole inactive nondegenerate level $0^L$, dipole active twofold degenerate level $1^U$ and dipole active nondegenerate level $0^U$. The efficiency of the exchange interaction and the magnitude of the splitting strongly decrease with increasing the QD size (approximately as $1/R^3$). As seen in the Fig. 3, in spherical CdS QDs with sizes about 2 nm, the splitting of states due to the electron-hole exchange interaction is smaller than 1 meV. Therefore, the effects due to the exchange interaction can be safely neglected when calculating, e.g., the absorption spectra for the ensembles of QDs with realistic size dispersion. However, as we have shown in Ref. 14, the exchange interaction is crucial when treating the spin relaxation in a QD.

To sum up, the lowest electron state is the *S*-state and, when the influence of the split-off bands is considered, the lowest hole state is the *P*-state. The different spatial envelope function symmetries of the electron and hole ground states seemingly imply that in CdS QDs the exciton ground state should be an optically passive "dark exciton"[11,32]. However, thanks to the downward shift of the exciton energy levels due to the Coulomb interaction, the lowest state of an exciton is of *S*-symmetry for a wide range of QD sizes and also for different sets of material parameters. When the electron-hole exchange interaction is considered the lowest exciton state $(1S, 1S_{3/2})_{s_z,F_z}$ (with $|F_z|$ = 3/2) is split to dipole inactive level with $|N_z| = 2$ (exciton ground state) and dipole active level $1^L$. Our results are in line with those obtained recently for CdS QDs by ab-initio calculations[13].

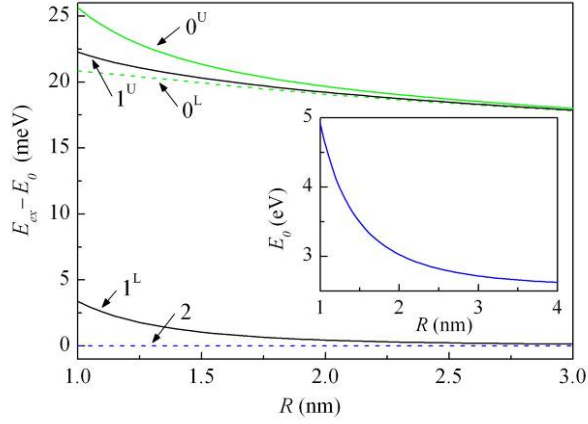

FIG. 3: Exchange field splitting of the lowest exciton state $(1S,1S_{3/2})_{s_z F_z}$ with $|s_z| = 1/2$ and $|F_z| = 1/2$ (upper group), and $|s_z| = 1/2$ and $|F_z| = 3/2$ (lower group) in CdS spherical QDs of different radii $R$. Energies are labeled by the values of $|N_z|$, where $N_z$ is the $z$-projection of the total angular momentum of the exciton ($N_z = s_z + F_z$); the index $U$ (upper) and $L$ (lower) is used to distinguish the states with the same value of $|N_z|$. The energies are counted from the lowest energy level of an electron-hole pair in QDs, which is shown in the inset. Dipole active and inactive levels are shown by solid and dashed lines, respectively.

### D. Connections with polarization-sensitive experimental results

The computed exciton fine structure is directly connected with the polarization-sensitive pump-probe experiments described in the following section. The pump-induced change in the transmitted intensity of the probe pulse can be computed from the occupancy factors of the energy levels as

$$\Delta I(\mathbf{e}_{\text{probe}}) \propto \int_0^\infty d\Omega\, I_{\text{probe}}(\Omega) \int_0^\infty dR\, \mathsf{N}(R) \frac{1}{4\pi} \int_0^\pi d\theta\, \sin\theta \int_0^{2\pi} d\varphi \sum_{s_z, F_z} \left| d^{(ex)}_{s_z F_z}(\mathbf{e}_{\text{probe}}) \right|^2 \\ \times w_{s_z F_z}(\mathbf{e}_{\text{pump}}) \delta(\hbar\Omega - E_{F_z}), \quad (13)$$

where $I_{\text{probe}}(\Omega)$ is the spectral distribution of the probe-pulse intensity, the function $\mathsf{N}(R)$ describes the distribution of QDs over radii, $d^{(ex)}_{s_z F_z}(\mathbf{e}_{\text{probe}})$ is the dipole matrix element of a transition between the exciton vacuum state and the exciton state $(1S,1S_{3/2})_{s_z F_z}$, $\mathbf{e}_{\text{pump}}$ and $\mathbf{e}_{\text{probe}}$ describe polarization of the pump and probe pulses, respectively, $w_{s_z F_z}(\mathbf{e}_{\text{pump}})$ describes the "blocking effect" of an exciton, created by the pump pulse in the same QD, on the absorption of probe light to the exciton state with given $s_z$ and $F_z$:

$$w_{s_z F_z}(\mathbf{e}_{\text{pump}}) = \sum_{s'_z F'_z} f_{s'_z F'_z}(\mathbf{e}_{\text{pump}}) \left( \delta_{s_z s'_z} + \delta_{F_z F'_z} - \delta_{s_z s'_z} \delta_{F_z F'_z} \right). \quad (14)$$

$f_{s_z F_z}(\mathbf{e}_{\text{pump}})$ is proportional to the probability that the exciton state with given $s_z$ and $F_z$ is populated as a result of pumping. In the absence of relaxation one has:

$$f_{s_z F_z}(\mathbf{e}_{\text{pump}}) = \int_0^\infty d\Omega \, I_{\text{pump}}(\Omega) \left| d_{s_z F_z}^{(ex)}(\mathbf{e}_{\text{pump}}) \right|^2 \delta(\hbar\Omega - E_{F_z}), \tag{15}$$

where $I_{\text{pump}}(\Omega)$ is the spectral distribution of the pump-pulse intensity. When writing down Eq. (13) we neglect the effects due to the (relatively weak) exciton-exciton interaction.

The degree of circular polarization of the signal $P_C$, which is determined experimentally in the pump-probe experiment (see Eq. (1a)), is given by the expression

$$P_C = \frac{\Delta I(\mathbf{e}_{\text{probe}}^+)/I(\mathbf{e}_{\text{probe}}^+) - \Delta I(\mathbf{e}_{\text{probe}}^-)/I(\mathbf{e}_{\text{probe}}^-)}{\Delta I(\mathbf{e}_{\text{probe}}^+)/I(\mathbf{e}_{\text{probe}}^+) + \Delta I(\mathbf{e}_{\text{probe}}^-)/I(\mathbf{e}_{\text{probe}}^-)}, \tag{16}$$

where $\Delta I(\mathbf{e}_{\text{probe}}^+)$ [$\Delta I(\mathbf{e}_{\text{probe}}^-)$] is the pump-induced change of the intensity of co-polarized (oppositely polarized) probe pulses with the incoming intensity $I(\mathbf{e}_{\text{probe}}^+)$ [$I(\mathbf{e}_{\text{probe}}^-)$]. Similarly, the degree of linear polarization of the signal $P_L$ (see Eq. (1b)) is given by the expression

$$P_L = \frac{\Delta I(\mathbf{e}_{\text{probe}}^\parallel)/I(\mathbf{e}_{\text{probe}}^\parallel) - \Delta I(\mathbf{e}_{\text{probe}}^\perp)/I(\mathbf{e}_{\text{probe}}^\perp)}{\Delta I(\mathbf{e}_{\text{probe}}^\parallel)/I(\mathbf{e}_{\text{probe}}^\parallel) + \Delta I(\mathbf{e}_{\text{probe}}^\perp)/I(\mathbf{e}_{\text{probe}}^\perp)}, \tag{17}$$

where the superscript $\parallel$ ($\perp$) corresponds to the case when the polarization vector of probe pulses is parallel (perpendicular) to that of pump pulses.

The transition probability in spherical QDs is the same for a pair of states, which differ from each other only by the sign of $N_z$. This means that the absorption of the linearly polarized light produces the same number of the excitons with the spin-up and spin-down orientation (i.e., parallel and anti-parallel with the k-vector of light). For this reason, measurements with linearly polarized light are not useful for studying the electron spin dynamics in the spherical QDs. Still, due to the presence of lattice (or shape) anisotropy in the assembly of QDs with a hexagonal crystal structure, it is possible to excite the non-zero linear polarization. Dynamics of $P_C$ and $P_L$ in the model are determined by relaxation processes, which modify the occupation probabilities $f_{s_z F_z}(\mathbf{e}_{\text{pump}})$ for exciton states with a given $N_z = s_z + F_z$. The "initial values" of these probabilities – i.e. the values in the absence of relaxation – are given by Eq. (15). The values of $P_L$ are determined by different pump-induced occupations of states with different $|N_z|$ (e.g., $N_z = 1$ and $N_z = 0$). The $P_L$ dynamics is thus mainly related to the transitions between states with different $|N_z|$. On the other hand, the light pulses with a fixed photon energy and opposite circular polarizations are absorbed into the states with opposite $N_z$ (e.g., $N_z = +1$ and $N_z = -1$), i.e. the states with the opposite exciton spin projection. The values of $P_C$ and their dynamics are determined by the different pump-induced occupations of the states with different signs of $N_z$ (with the same $|N_z|$) and to the relaxation between these states. So, just the measurements of $P_C$ are relevant to the spin dynamics in spherical QDs. From this reason we will further concentrate more on dynamics of $P_C$ than on that of $P_L$.

### E. Dynamics and spectral dependence of $P_C$ and $P_L$

The decay of $P_C$ is widely used for the examination of the electron spin relaxation. However, as we will show below, the $P_C$ decay may be influenced also by relaxation processes where the electron spin is preserved. In the following, we present the dynamics of the circular polarization of the signal $P_C$ calculated from the occupation probabilities $f_{s_z F_z}(\mathbf{e}_{\text{pump}})$. The occupation probabilities are strongly affected by the interaction of the excitons with phonons. Because the energy distance between the exciton states is different from the optical-phonon energies in CdS, the one-phonon processes, which could modify the occupation probabilities of the exciton levels, have negligible amplitude. However, according to our calculations, at room temperatures quasi-elastic two-LO-phonon processes without electron spin flip[21] may give an important contribution to the dynamics of $P_C$ on *picosecond* time scales. In those transitions, one LO-phonon is absorbed and another LO-phonon is emitted (so that the energies of the initial and final exciton states must be close to each other). As only the hole angular momentum $F_z$ can be changed by the phonon interaction, the initial and the final states have different absolute value of $|N_z|$. Because these transitions take place between different levels (see. Fig. 3), we call these transitions *interlevel* transitions. Rates of the above transitions are calculated according to Fermi's golden rule and the corresponding time constants reach values of the order of picoseconds (see Fig. 5 in Ref. 21). These transitions reduce the magnitude of $P_C$, but not to zero; in equilibrium the $P_C$ reaches 1/3 of its initial value. Note that the electron spin is not affected by these transitions. Further relaxation is possible between the exciton states with opposite z-projections of the total exciton angular momentum, $N_z = 1$ and $N_z = -1$, which includes the electron spin flip. These LO-phonon-assisted *intralevel* transitions require electron spin flip, which is possible due to the (relatively weak) electron-hole exchange interaction. Just the fact that the electron-hole exchange interaction is weak leads to relatively low values of the corresponding transition rates. Rates of electron-spin-flip transitions between degenerate exciton states with $|N_z| = 1$ are calculated according to the Fermi golden rule to the lowest order in the electron-hole exchange interaction. The corresponding characteristic time of the $P_C$ decay due to this mechanism is of the order of several nanoseconds[14].

Our model implies that the dynamics of $P_C$ will contain two components with substantially different time constants (as shown in Fig. 4), where the slower one is connected to the electron spin relaxation. Our calculations were done for an ensemble of randomly oriented CdS QDs with a hexagonal crystal structure and a realistic size distribution. The bottom curve in Fig. 4 corresponds to the case when the wavelength of laser pulses (photon energy $h\nu$) is tuned to resonance with the lowest optically active energy level in a typical QD from the distribution of QDs sizes ($E_1$ in the following). However, for this photon energy not only the lowest optically active state in a typical QD is investigated – also higher lying (optically active) states in larger QDs are probed simultaneously in a realistic (inhomogeneously broadened) ensemble of QDs. When the photon energy is decreased (i.e., for negative values of the detuning energy of photons $\Delta E = h\nu - E_1$) only lower lying states in larger QDs are excited, which corresponds to the regime of a size-selective excitation[5,11,33]. In the inset of Fig. 4 we show the calculated values of $P_C$ and $P_L$ after the "fast" (electron-spin conserving) exciton relaxation as a function of the detuning energy $\Delta E$. Due to the simultaneous excitation of different exciton states in QDs of different sizes, the values of $P_C$ and $P_L$ are non-monotonous functions of the detuning energy $\Delta E$. In particular, the values of $P_C$ are increasing for more negative values of $\Delta E$ due to the relative enhancement of the absorption to the lowest dipole-active exciton states $(1S,1S_{3/2})_{\mp 1/2,\pm 3/2}$ (i.e., states with $|F_z|=3/2$ and $|N_z|=1$ in Fig. 3). Resonant absorption of a circularly polarized light into the lowest $(1S,1S_{3/2})_{S_z,F_z}$ exciton state of one QD should, in principle, produce a circular polarization of the population $P_N = 100\%$. However, in the assembly of randomly oriented QDs with a hexagonal crystal structure, the averaging lowers the maximum achievable spin polarization to $P_N = 5/7 = 81\%$ (Ref. 34). The signal polarization in our particular case is further lowered by the averaging through the spectral width of the ultrashort laser pulses and through the inhomogeneously broadened distribution of QDs sizes.

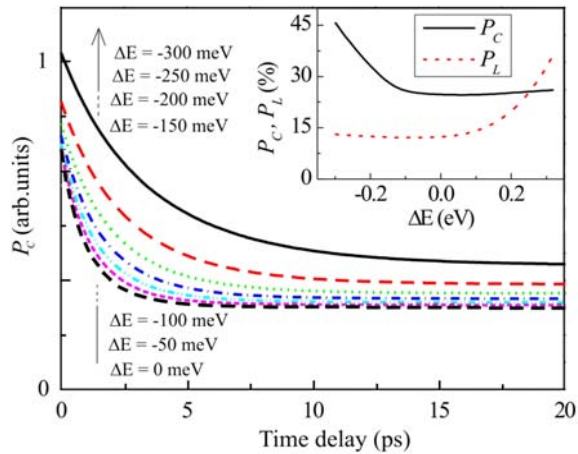

FIG. 4: Dynamics of $P_C$ calculated for an ensemble of CdS spherical QDs (characterized by a lognormal size distribution with $R_0 = 2.1$ nm and $\rho = 0.25$) for different values of the detuning energy $\Delta E$ ($\Delta E = h\nu - E_1$; $h\nu$ is the photon energy and $E_1$ is energy the lowest optically active state in a typical QD from the distribution of QDs sizes). Inset: Calculated values of $P_C$ and $P_L$ after the "fast" (electron-spin conserving) exciton relaxation as a function of $\Delta E$. Calculations are performed for the spectral width of laser pulses of 12 meV.

## IV. EXPERIMENTAL RESULTS AND DISCUSSION

### A. Polarization-insensitive measurements

We investigated samples with a different mean size of QDs produced by three different manufacturers (Hoya, Schott and Corning). In Fig. 5 (a) we show the room temperature absorption spectra of the studied samples. There are several quantized (inhomogeneously broadened) transitions in the spectra, which is a feature typical for CdS QDs in glass matrix[1]. The spectral positions of the transition energies (i.e., the positions of the absorption peaks) can be determined from the second derivative of the measured absorption spectra where they are apparent as rather pronounced minima[35]. The positions of the lowest transition energy ($E_1$), which corresponds to the lowest optically active state in a "typical" QD (i.e., in a QD with the most probable size within the corresponding size distribution), are shown as arrows in Fig. 5 (a) for all the samples. The radius $R_0$ of a "typical" QD in a given sample can be determined from $E_1$ using the theoretical dependence of the lowest transition energy shown in the inset in Fig. 3 - the obtained values of $R_0$ are 1.9, 2.0, 2.1, and 2.7 nm for samples GG435, 3-73, Y44, and Y46, respectively.

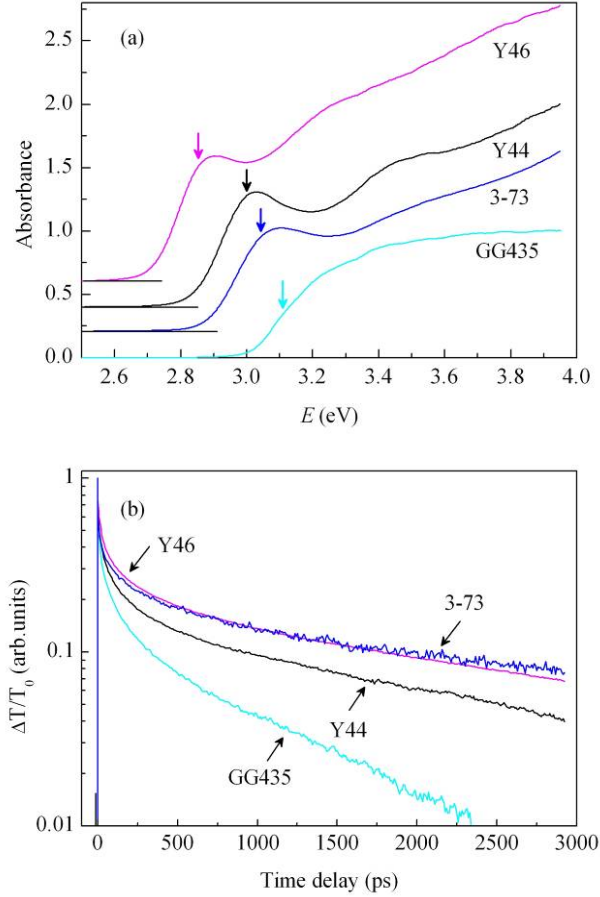

FIG. 5: (a) Absorption spectra of the investigated samples at 300 K. Vertical arrows depict the positions of the lowest transition energy $E_1$ in the samples, which were determined from the second derivative of the absorption spectra. The three upper curves were vertically shifted for clarity. (b) Polarization-insensitive dynamics of differential transmission $\Delta T/T_0$ measured at 300 K with the photon energy tuned to $E_1$ for each sample; the data are normalized.

Polarization-insensitive dynamics of differential transmission $\Delta T/T_0$, measured with the photon energy tuned to $E_1$ for each of the samples, are shown in Fig. 5 (b). The observed dynamics, which reflect the energy relaxation and recombination of photoinjected carriers, are not monoexponential, but rather consist of a fast initial decay (components with a characteristic time constants from tens of femtoseconds to several picoseconds) followed by a slower (nanosecond) decay - note a logarithmic scale in the figure. These results are quite typical for II-VI QDs in glass matrix and can be explained as a consequence of a fast trapping of a part of the carriers to the QDs surface states[36] and a recombination. The trapped carriers do not contribute to the $\Delta T/T$ signal (because they are not in the energy window that is monitored by probe laser pulses) but they can eventually lead to the existence of the trap-related photoluminescence[37]. The long component of the $\Delta T/T$ decay reflects the recombination time of the carriers, which stayed in the energy states around $E_1$ in the volume of nanocrystals. However, we have not observed any systematic dependence of the dynamics of $\Delta T/T_0$ on $R_0$ [e.g., very similar curves were detected for samples Y46 and 3-73 with rather different values of $R_0$ – see Fig. 5 (b)]. This clearly shows that *not only* the QDs size is important for the carrier dynamics. For example, the passivation of the surface of individual QDs (Ref. 36) and/or a chemical composition of the host glass[20] can play an important role. All these sample-dependent properties significantly complicate any quantitative analysis of the obtained dependences on QDs size, which are measured in *different samples*. Instead, it might be much more convenient to measure the desired size dependences in *one sample*. This can be done by a resonant laser excitation of a certain fraction of QDs from the whole distribution of QDs sizes. For example, QDs with radius from 1 to more than 3 nm are present in sample Y44 - see Fig. 1 (b) in Ref. 21. And it is much easier to compare the data obtained in such a manner with the predictions of the theoretical model provided that the relevant distribution of QDs sizes is known. In Fig. 6 (a) we compare the measured absorption spectrum in sample Y44 (open points) with the calculated one (dashed line). The calculation of the absorption into the first $(1S, 1S_{3/2})$ state for the ensemble of randomly oriented spherical CdS QDs with a hexagonal crystal structure was performed in a dipole approximation, neglecting the exchange interaction splitting, which is small compared to the inhomogeneous broadening (see Fig. 3). The deduced inhomogeneous width of the size distribution $\rho^{CALC} = 0.1$, which best fits the measured absorption spectrum, is smaller than the one determined from the analysis of TEM images $\rho^{TEM} = (0.27 \pm 0.02)$. This might indicate that the size distribution function for the whole sample is actually narrower than that for the subset of QDs observed in HRTEM images. Nevertheless, it is clear from Fig. 6 (a) that our model for this particular sample is reasonably valid only up to the energy ~ 3.1 eV. For higher energies, excited electron states, which are not considered in our model, contribute significantly to the absorption.

The absorption spectrum of the nanocrystalline sample is inhomogeneously broadened due to the size distribution of the nanocrystals. In Fig. 6 (b) we show the initial stages of carrier dynamics measured across this inhomogeneously broadened absorption spectrum of the sample Y44. In this experiment, the spectral resolution is given by the spectral width of the femtosecond laser pulses, which is schematically depicted in the lower part of Fig. 6 (a). The obtained normalized dynamics of $\Delta T/T_0$ clearly show an acceleration of carrier dynamics when increasing the detuning energy $\Delta E$. This behavior reflects a decrease of QDs sizes, which are sampled by the laser pulses, and the fact that not only the lowest energy states of the smaller QDs are examined for larger photon energy, but also the higher energy states of the larger QDs can be excited at the same time. In fact, the former effect is probably dominant below $\Delta E \approx 100$ meV (in smaller QDs the influence of carrier trapping to the QDs surface states is stronger) while above this value the latter effect prevails. In Fig. 6 (a) we show the spectral dependence of the initial magnitudes of $\Delta T/T_0$ that were measured in sample Y44 for time delay of 0.25 ps (solid points) together with those that were calculated from exciton energy levels shown in Fig. 3 (solid line in Fig. 6 (a)). The differential transmission $\Delta T/T_0$ is positive in the whole investigated spectral region reflecting the state-filling induced saturation of the exciton levels. The measured spectral dependence of $\Delta T/T_0$ has a maximum at the position of the transition energy $E_1$ (when a maximal number of QDs from the whole distribution of QDs sizes was excited) and decreases for other photon energies. The experimental spectral dependence is narrower than the calculated one, possibly due to the admixture of the biexciton-related induced absorption[38,39,40], which was not included in our computation.

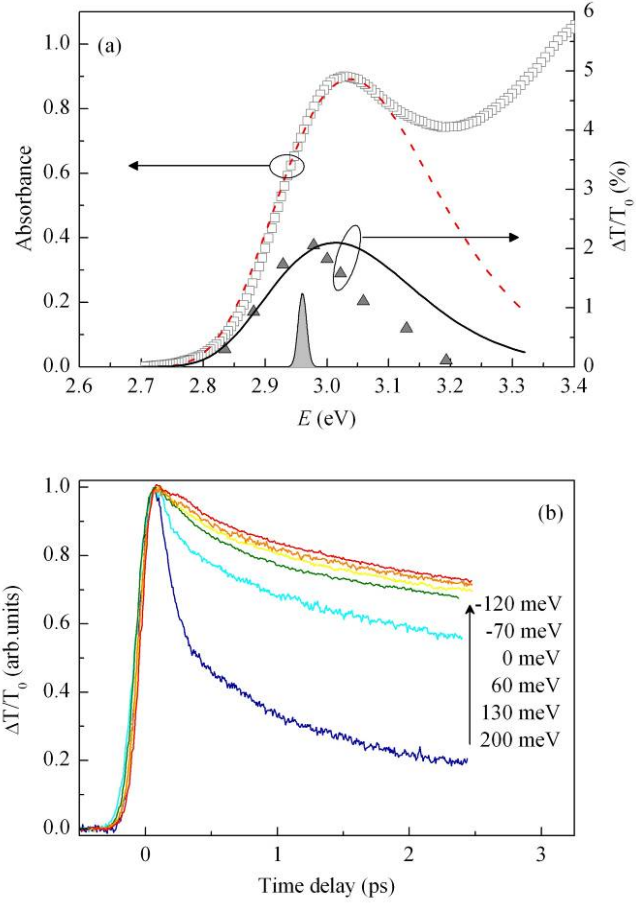

FIG. 6: (a) Absorption spectrum (open squares) and spectrum of differential transmission $\Delta T/T_0$ (at time delay of 0.25 ps; full triangles) measured in sample Y44 at 300 K. Theoretical absorption spectrum (dashed line) and spectrum of $\Delta T/T_0$ (solid line) were calculated for an ensemble of QDs characterized by a lognormal size distribution with $R_0 = 2.05$ nm and $\rho = 0.1$ and for the spectral width of laser pulses of 12 meV. (b) Polarization-insensitive dynamics of $\Delta T/T_0$ measured across the inhomogeneously broadened absorption spectrum of sample Y44 at 300 K with various detuning energies $\Delta E$; the data are normalized. The spectral resolution of the experiment is given by the spectral width of the femtosecond laser pulses, which is schematically depicted in the lower part of panel (a).

One can see that our theoretical model provides a qualitative description of all the features observed experimentally. The major problem for the quantitative comparison between the experimental results and the theoretical predictions is the fast carrier trapping to the surface of some individual QDs, which is generally believed to be the dominant recombination channel in these small QDs (Ref. 36). And even though we successfully reduced the influence of the sample-dependent properties (e.g., the variation of the host glass between the different samples) by measuring the desired QDs size dependences in one sample, the absence of an accurate theoretical treatment of the surface trapping precludes any direct comparison of the measured and computed *polarization-insensitive* dynamics of $\Delta T/T_0$. On the other hand, the *polarization-sensitive* dynamics, which will be treated in the next section, *does not* suffer from this problem because the dynamics of $P_C$ and $P_L$ are already corrected for the finite lifetime of carriers[14].

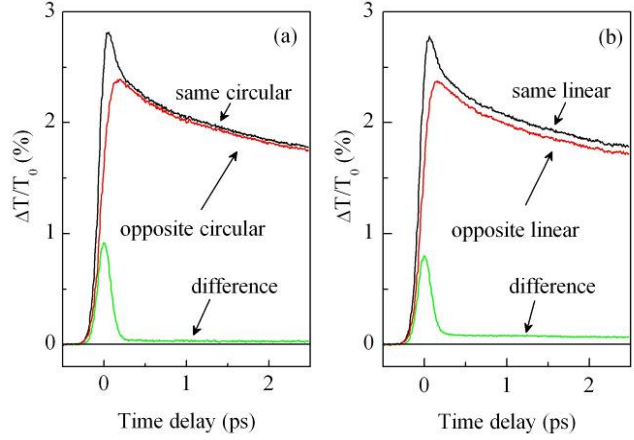

FIG. 7: (a) Polarization-sensitive dynamics of $\Delta T/T_0$ measured in sample Y44 at 300 K using probe pulses with the same and opposite circular polarization with respect to the circular polarization of the pump pulses. The lower curve is the signal difference that was measured directly using the photoelastic modulator. (b) The same as (a) but with linearly polarized light. The photon energy was tuned to the spectral positions of $E_1$.

## B. Polarization-sensitive measurements

Polarization-sensitive dynamics of $\Delta T/T_0$ measured by circularly and linearly polarized laser pulses are shown in Fig. 7 (a) and Fig. 7 (b), respectively. There is a clear (though rather weak) dependence of the signal on the probe polarization. The difference signal, which was directly measured using a lock-in amplifier and a photo-elastic modulator (PEM), is shown in the lower part of the figure. The spectral dependence of the dynamics of $P_C$ [$P_L$] measured at 300 K and 8 K is shown in Fig. 8 (a) and (b) [(c) and (d)], respectively. The most pronounced feature, which was observed in the dynamics of $P_C$ and $P_L$ for all samples and all temperatures investigated, is that the signals decay at two rather distinct time scales: the values decrease first with a sub-picosecond time constant and then the decay is considerably slower. The measured experimental data can now be compared with the predictions of our model (see Fig. 4). And even though there is not a perfect agreement between the experimental and theoretical results, the major features observed in the experiment, namely, the existence of the two component decay and a strong enhancement of $P_C$ for negative values of $\Delta E$, are reproduced well by the theory. Therefore, we attribute the "fast" component in the experimentally observed decay to the electron-spin-conserving relaxation of the energy and total angular momentum of holes (*interlevel* exciton transitions) while the "slow" component is attributed to *intralevel* exciton transitions that require electron spin flip[14]. Note also that when the size selective excitation regime is reached (i.e., for large negative values of $\Delta E$) strongly damped oscillations in both $P_C$ and $P_L$ are observed experimentally short after the excitation. These oscillations can be interpreted as coherent quantum beats between the individual QD states[41], which are strongly damped due to the inhomogeneous broadening in the sample. The initial stages of the measured dynamics, where the pump and probe pulses overlap in time (this region is shaded in Fig. 8), can be, in principle, affected also by the coherent scattering of pump and probe pulses that complicates any quantitative analysis of the "fast" component.

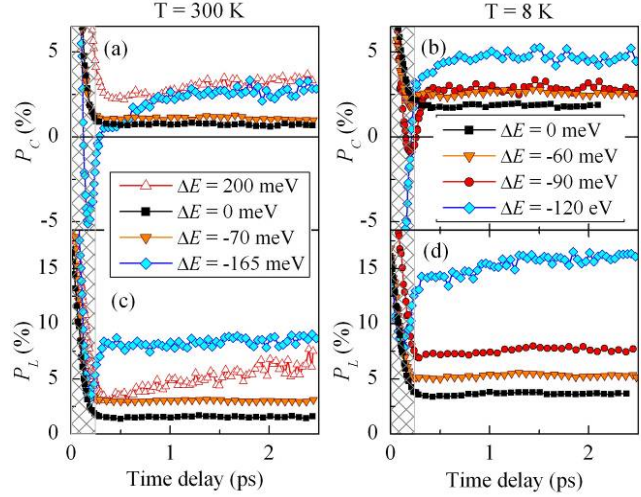

FIG. 8: Dynamics of $P_C$ [panels (a) and (b)] and $P_L$ [panels (c) and (d)] measured at 300 K (left panels) and 8 K (right panels) in sample Y44 for different detuning energies $\Delta E$. The shaded areas show schematically the temporal regime where the coherent scattering of pump and probe pulses can take place.

In the following, we will concentrate on the "slow" component of the decay of $P_C$ and $P_L$. In Fig. 9 (a) and Fig. 9 (b) we show the spectral dependence of the magnitude of $P_C$ (full points) and $P_L$ (open points) measured at the time delay of 2 ps in sample Y44 at 300 K and 8 K, respectively. (At 8 K we were not able to measure the data for $\Delta E > 0$ due to the temperature-induced shift of the absorption spectrum[37] out of the spectral tuneability of our laser system.) The data obtained at 300 K in samples 3-73 and Y46 are shown in Fig. 9 (c) and Fig. 9 (d), respectively. While the details of the measured dependencies are sample dependent, there are certain features that are common for all the samples. First, the values of $P_C$ and $P_L$ are minimal around $\Delta E = 0$ and they increase both for larger and smaller values of $\Delta E$. Second, the magnitude of $P_L$ is always slightly larger than that of $P_C$. For negative values of $\Delta E$, where our theoretical model is applicable (cf. Fig. 6 (a)), the measured spectral dependence of $P_C$ and $P_L$ (Fig. 9) can be compared with the theoretical predictions (see the inset in Fig. 4). As already discussed above, both in experiment and theory we see an increase of $P_C$ for negative values of $\Delta E$. However, the measured values of $P_C$ are considerably smaller than the computed ones and also the ratio $P_L / P_C$ is underestimated by our model. Both these effects are probably connected with the shape anisotropy in our samples[34], which is a consequence of the slightly non-spherical shape of QDs in the investigated samples [cf. Fig. 1 (a) in Ref. 21]. This explanation is also supported by the comparison of our experimental results with the polarization-sensitive measurements reported for the self-assembled QDs. The strongly non-spherical shape of these self-assembled QDs (Ref. 42) leads to drastic enhancement of $P_L$ ($\approx 80\%$) relative to $P_C$ (< 5 %) that is typically measured for undoped samples without external magnetic field[43]. Alternatively, fast trapping of electrons to surface states in some QDs may reduce to zero the contribution of those QDs to the "difference signal", $\Delta T / T_0^{++} - \Delta T / T_0^{+-}$, while the "sum signal", $\Delta T / T_0^{++} + \Delta T / T_0^{+-}$, may remain almost the same (provided that holes are not trapped). If such a trapping of electrons occurs in an appreciably large fraction of QDs, the value of $P_C$ measured by the probe pulse could significantly decrease.

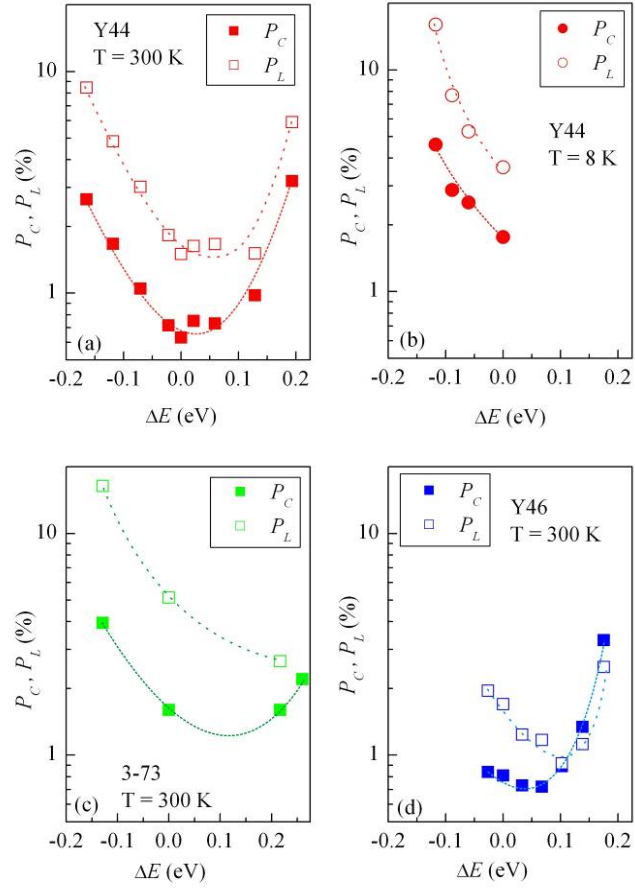

FIG. 9: Spectral dependence of $P_C$ (full points) and $P_L$ (open points) measured at time delay of 2 ps in sample Y44 at 300 K (a) and 8 K (b), sample 3-73 at 300 K (c), and sample Y46 at 300 K (d). Lines are guides to the eye.

Finally, we consider the room temperature spin-related exciton dynamics, which is connected with the nanosecond decay of $P_C$. In our previous paper we have shown that the spin relaxation is faster in smaller QDs (due to the increase of the efficiency of the electron-hole exchange interaction, which depends on the spatial overlap of their wavefunctions)[14]. The experimental values of the electron spin relaxation rate [$\Gamma = 1/(2\tau)$, where $\tau$ is the time constant that describes the long component of the $P_C$ decay[14]], which were reported previously, were measured entirely with the photon energy tuned to the spectral positions of $E_1$ in different samples[14]. In Fig. 10 we show that the observed size dependence of $\Gamma$ is maintained also when the laser pulses are tuned across the inhomogeneously broadened absorption spectra of the investigated samples. This further confirms that the size dependent properties of QDs can be effectively studied also in one sample by the resonant laser excitation of a certain fraction of QDs from the whole distribution of QDs sizes. The lines in Fig. 10 are the transition rates between the lowest dipole-active exciton states $(1S,1S_{3/2})_{-1/2,3/2}$ and $(1S,1S_{3/2})_{1/2,-3/2}$ that were calculated in the framework of the electron spin relaxation mechanism, suggested in our previous paper (namely, two-LO-phonon intralevel transitions with the electron spin flip, driven by the electron-hole exchange interaction[14]). We performed the calculations with three different sets of effective-mass parameters: $\gamma_1 = 1.71$, $\gamma_2 = 0.62$, $\Delta_{so} = 70$ meV (Ref. 29); $\gamma_1 = 1.02$, $\gamma_2 = 0.41$, $\Delta_{so} = 62.4$ meV (Ref. 30); $\gamma_1 = 1.09$, $\gamma_2 = 0.34$, $\Delta_{so} = 68$ meV (Ref. 7). Note that no fitting parameters are used in our calculations[14]. From the comparison of the experimental and theoretical results shown in Fig. 10 it follows that the set of effective-mass parameters reported in Ref. 30 (set B) does not seem to describe well the investigated CdS QDs in a glass matrix. The reasons for this discrepancy can be qualitatively understood from the corresponding exciton energy spectra, shown in Fig. 2 and determined by the parameters $\gamma_1$, $\gamma_2$, and $\Delta_{SO}$, which are different for sets A, B, and C. As compared to sets A and C, for set B the energy spectrum contains excited states, separated from the states $(1S,1S_{3/2})_{-1/2,3/2}$ and $(1S,1S_{3/2})_{1/2,-3/2}$ by energy intervals, which are closer to the LO phonon energy (about 38 meV). For those "better matching" energy intervals, the two-phonon assisted spin relaxation processes[14] through (intermediate) excited states are more efficient and hence the spin relaxation rate is significantly enhanced. The other two sets of parameters provide similar results and spin relaxation rates computed with the parameter set reported in Ref. 7 seems to be the most similar to the experimentally obtained values. Our main point is that an adequate description of the measured dynamics requires an accurate description of all relevant exciton states. As implied by Fig. 10, the calculated relaxation rates, corresponding to two-LO-phonon intralevel transitions with the electron spin flip driven by the electron-hole exchange interaction[14], are strongly affected by a specific pattern of the lowest exciton energy levels, including, of course, also a specific sequence of the exciton energy levels. In other words, the relevance of the used model is determined not only by the derived sequence of the ($1S_e$, $1P_h$) and ($1S_e$, $1S_h$) exciton energy levels but by the whole set of lowest exciton states in the energy range of the order of few LO-phonon energies.

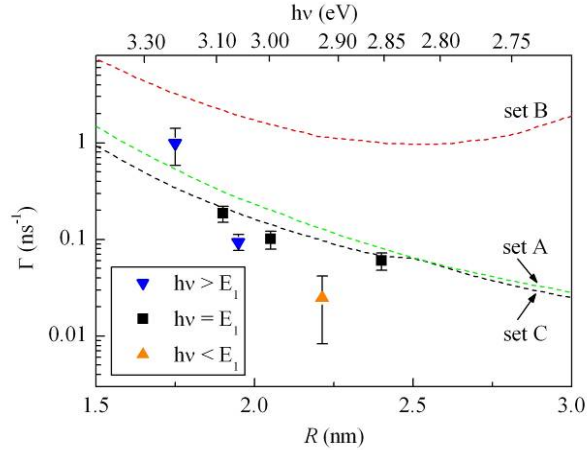

FIG. 10: Room temperature spin relaxation rate $\Gamma$ as a function of the QD radius $R$. The experimental values (points) were measured across the inhomogeneously broadened absorption spectra of the samples using laser pulses with different values of photon energy $h\upsilon$. The curves show the rate of transitions between the lowest dipole-active exciton states $(1S,1S_{3/2})_{-1/2,3/2}$ and $(1S,1S_{3/2})_{1/2,-3/2}$ that was calculated for 3 sets of parameters: set A ($\gamma_1 = 1.71$, $\gamma_2 = 0.62$, $\Delta_{SO} = 70$ meV) (Ref. 29), set B ($\gamma_1 = 1.02$, $\gamma_2 = 0.41$, $\Delta_{SO} = 62.4$ meV) (Ref. 30), and set C ($\gamma_1 = 1.09$, $\gamma_2 = 0.34$, $\Delta_{SO} = 68$ meV) (Ref. 7).

## V. CONCLUSIONS

We calculated the exciton fine structure for CdS spherical QDs with the wurtzite crystal structure. If the influence of the split-off bands is taken into account, the lowest *hole* state is a *P*-state for a wide range of QD sizes and also for different sets of material parameters.

However, due to the electron-hole Coulomb interaction, the lowest state of an *exciton* appears to be of *S*-symmetry both for the electron and hole components. This state is further split by the crystal field and the electron-hole exchange interaction into a dipole inactive level with $|N_z|=2$, which is a ground state of the exciton, and dipole active level $1^L$. This implies that the electron-hole exchange splitting rather than the spatial symmetry may explain the nature of the "dark exciton" in CdS QDs. Based on this calculated structure of exciton energy levels we constructed a model for the exciton spin dynamics in spherical QDs. In particular, we computed the spectrally-resolved polarization-sensitive dynamics of a differential transmission in an ensemble of randomly oriented CdS QDs with a realistic size distribution, which can be directly compared with our results that were measured in CdS QDs in glass matrix. The qualitative agreement between the experimental results and theoretical predictions enabled us to assign the "fast" component observed in the decay of $P_C$ to the electron-spin-conserving relaxation of the energy and total angular momentum of holes. The "slow" component in the decay of $P_C$ was attributed to the relaxation between the (quasi-)degenerate exciton states with opposite *z*-projections of the total exciton angular momentum, $N_z=1$ and $N_z=-1$, which involves the electron spin flip. The obtained dependence of the electron spin relaxation rate on the radius of QDs, which was both measured and computed, enabled us to select the set of effective-mass parameters that provide the best description of CdS QDs in a glass matrix.

**ACKNOWLEDGMENTS**

This work was supported by Ministry of Education of the Czech Republic in the framework of the research center LC510 and the research plan MSM0021620834, by the Fund for Scientific Research - Flanders Projects G.036508N, G.011506, G.0356.06, by the WOG WO.035.04N (Belgium), and the EC Network of Excellence SANDiE, Contract No. NMP4-CT-2004-500101. P.H., neé Nahálková, acknowledges the support of the Special Research Fund of the University of Antwerp, BOF NOI UA 2004